\documentclass[prl,twocolumn,showpacs,preprintnumbers,superscriptaddress]{revtex4-1}

\usepackage[utf8]{inputenc}
\usepackage{amssymb}
\usepackage{amsmath}
\usepackage{graphicx}
\usepackage{subfigure}
\usepackage{dcolumn}
\usepackage{bm}
\usepackage{hyperref}
\usepackage{enumerate}
\usepackage{xcolor}
\usepackage{natbib}
\usepackage{amssymb}
\usepackage{amsbsy}
\usepackage{graphics}
\usepackage{comment}
\usepackage{hyperref}

\begin{document}

\preprint{APS/123-QED}

\title{Hierarchical Jamming in Frictional Particle Assemblies}

\author{R. Li}
\author{D. R. Lester}
\affiliation{School of Engineering, RMIT University, 3000 Melbourne, Australia}
\email{daniel.lester@rmit.edu.au}

\date{\today}

\begin{abstract}
The postulate of the existence of a jamming phase diagram (Liu and Nagel, Nature 396, 21 EP (1998aa)) provides a theoretical basis for the classification of a wide range of amorphous solids (colloidal, molecular and emulsion glasses, colloidal and polymer gels, foams and granular matter) on the basis of whether these materials are in the jammed or unjammed state. Whilst such simple classification is appealing, it fails to capture that the criterion of \emph{rigidity} of such amorphous solids may be defined with respect to a particular deformation orientation or mode (i.e. shear, extrusion, consolidation). We consider this problem via the consolidation of strong colloidal gels, and find that the critical transitions during the consolidation of a strong colloidal gel (as indicated by maxima and minima in the relative normal stress difference) correspond directly to directed, frictional and non-frictional rigidity percolation. These results indicate a hierarchy of directed, jammed states during consolidation of such amorphous solids, and a direct link between particle-scale interactions and macroscopic collective behaviour of these systems driven far from equilibrium.

\end{abstract}


\maketitle


The deformation and flow of soft matter - including colloidal, molecular and emulsion glasses, colloidal and polymer gels, foams and granular matter - is central to a vast array of man-made and natural applications. Typically, the mechanical behaviour of these complex fluids and amorphous solids is very complicated~\cite{Bonn:2009aa, Lu:2008aa, Mari:2014aa}. Such soft materials are characterised by an absence of long-range order at the microscale~\cite{Boromand:2018aa}, hysteresis under reversal of applied stress or strains~\cite{Parneix:2009aa}, non-local behaviour and critical transition(s) from fluid to solid-like behaviour~\cite{Lu:2008aa, Mari:2014aa}. Often spatial heterogeneity and non-local effects dominate near these non-equilibrium transitions, invalidating e.g. mean-field theories of macroscopic behaviour~\cite{Siemens:2010aa, Wyart:2008aa}.

In a bid to identify and characterize unified classes of behaviours across the diverse set of soft matter, these critical rigidity transitions have been described as \emph{jammming} transitions. Liu and Nagal~\cite{Liu:1998aa} postulate a jamming phase diagram (in terms of scaled temperature, pressure and applied load) to describe the jamming of these broad classes of materials, which has been utilised to develop quantitative theories of macroscopic jamming and flow.

Such jamming is defined as the onset of macroscopic structural rigidity~\cite{Wyart:2008aa, Liu:1998aa}, such that the microstructure can support some finite stress without plastic deformation. Whilst the concept of a unique jammed state unifies the behaviour of diverse range of disordered solids within the thermodynamic phase space~\cite{Wyart:2008aa}, this concept of \emph{rigidity} is not unique. For example, a soft material in a given microscopic state may be rigid with respect to one mode of deformation (e.g. shear, extrusion, consolidation) and not another, and moreover the rigidity itself may be directed, regardless of whether the material is intrinsically anisotropic~\cite{Kohl:2016aa, Takeuchi:2007aa, Majmudar:2005aa}. Hence classification solely in terms of a single jamming transition is insufficient to fully characterise the rigidity of many disordered solids.

In this study we consider macroscopic deformation of assemblies of attractive, frictional particles and show that directed, hierarchical jamming directly controls evolution of both the material microstructure and macroscopic stresses. Here this amorphous solid undergoes several distinct, punctuated regimes during deformation, and these regimes are bounded by critical jamming transitions which are quantified as frictional and frictionless rigidity percolation thresholds~\cite{Stauffer:1985aa, Song:2010aa, Wang:2011aa}. This hierarchy of directed jammed states illustrates how microscopic properties organise collective particle behaviour, and provide a direct link with non-equilibrium granular thermodynamics~\cite{Edwards:1989aa, Makse:2002aa, Martiniani:2017aa, Song:2010aa, Wang:2011aa}.

In this study we consider the quasi-static uniaxial consolidation of a strongly flocculated colloidal gel~\cite{Channell:1997aa, Boromand:2017aa}, but we shall show that the critical properties of this system and the concepts encountered herein are relevant to the general deformation (shear, extrusion, mixed) of frictional soft matter (granular matter, frictional glasses, weak gels and metamaterials). 
Whilst there have been many studies of the consolidation of strongly flocculated colloidal gels~\cite{Buscall:1987aa, Miller:1996aa, Channell:1997aa, Gilabert:2008aa, Lu:2008aa, Parneix:2009aa, Seto:2013aa, Roy:2016aa, Roy:2016ab, Roy:2016ac}, only a handful~\cite{Seto:2013aa,Roy:2016aa} have attempted to link particle-scale properties to macroscopic rheology. Particle-scale forces typically consist of electrostatic and depletion central forces, along with frictional forces arising from inter-particle torsion, rolling and sliding. Under quasi-static consolidation, slow macroscopic motion of the consolidation boundaries (i.e. filtration piston, sedimentation interface), give rise to collective particle motion as particles rearrange into more concentrated (yet largely disordered) packing. As hydrodynamic drainage does not impact microstructure evolution in this regime~\cite{Brambilla:2011aa} and the strongly flocculated state render the system athermal (non-Brownian), this many-body problem is governed solely by particle-particle interactions, and so is generic to the mechanics of disordered solids and granular matter more generally. 

Sir Sam Edwards' theory~\cite{Edwards:1989aa} of the thermodynamics of granular materials - recently validated for ergodic systems~ \cite{Baule:2018aa, Martiniani:2017aa} - is based upon an analogy between the void space per particle and absolute temperature. Under this framework, the translational and rotational degrees of freedom (DOFs) of particles within the microstructure decrease with increasing solids concentration $\phi$, leading to a series of jamming transitions at critical values of $\phi$ when a relevant subset of these DOFs are exhausted. These microscopic DOFs are not only consumed during the consolidation process~\cite{Makse:2002aa}, but they also control evolution of the microstructure itself due to preferential particle displacements that arise from the hierarchy of frictional and normal forces present in the system. 
Hence, whilst there exist many possible pathways for colloidal gels to consolidate from dilute to random close-packed, the observed (maximum likelihood) pathway is strongly dependent upon the relative magnitudes of the inter-particle forces. This path selection mechanism and associated jamming events during consolidation uncover the relationships between particle-scale forces, microstructural evolution and macroscopic deformation. We postulate that such path selection follows a maximum entropy production principle (MEPP)~\cite{Jaynes:1957aa, Dewar:2003aa, Martyushev:2006aa}, and so this approach, in conjunction with granular thermodynamics~\cite{Edwards:1989aa}, provide a means to quantify and upscale such non-equilibrium processes.

To explore these links more deeply we use the DEM algorithm described by Seto et al~\cite{Seto:2013aa} to simulate the quasi-static uniaxial consolidation of strongly aggregated colloidal gels comprised of 2D rigid particles (discs) with unit radii that interact through central forces, sliding and rolling friction. As these particles are strongly flocculated, Brownian motion is neglected, along with particle/fluid inertia in the limit of slow consolidation speed. Brambilla et al~\cite{Brambilla:2011aa} show that the effects of hydrodynamic drainage may also be neglected in this regime, hence the suspending fluid may be considered inert and the system behaves as an athermal assembly of attractive, frictional particles. Pair-wise additive forces are initiated upon contact between particle $i$ and $j$, which are comprised of a normal (central) force $\mathbf{F}_N^{(i,j)}$, sliding frictional force $\mathbf{F}_S^{(i,j)}$ and rolling frictional moment ${\bf M}_R^{(i,j)}$ (see \cite{SM:2000}). These normal, sliding and rolling forces are characterised by an elasto-plastic Hookean law with respective spring constants $k_N$, $k_S$ and $k_R$ and plastic limits $|\mathbf{F}_N^{(i,j)}|\leqslant F_{N,c}$, $|\mathbf{F}_S^{(i,j)}|\leqslant F_{S,c}$, $|\mathbf{M}_R^{(i,j)}|\leqslant M_{R,c}$, 
such that the frictional sliding force or rolling torque are respectively set to zero if $F_{S,c}$ or $M_{R,c}$  are exceeded, but the inter-particle bond persists. Conversely, the inter-particle bond is broken if $F_{N,c}$ is exceeded, but bonds may be reformed upon particle contact. This ratchet-like frictional model means that frictional contacts are \emph{saturated}; i.e. all contacts involve sliding and rolling friction such that when the plastic limits $F_{S,c}$ or $M_{R,c}$ are exceeded the corresponding force or moment instantly resets to zero and then continues to increase. Conversely, stick-slip friction models~\cite{Henkes:2016aa} allow for fully mobile sliding or rolling motion once the relevant plastic limits are exceed, leading to \emph{unsaturated} frictional contacts. Whilst these (and many other) frictional models may differ in microscopic detail, they are unified under a single elasto-plastic frictional framework which is 
the origin of macroscopic elasto-plastic behaviour in dense colloidal gels, hysteresis under small shear and compressive strains, strain hardening under bulk consolidation~\cite{Buscall:1987aa} and cyclical shear strain~\cite{Doorn:2018aa}. 

\begin{figure}[htb]
\centering
          \includegraphics[width=\columnwidth]{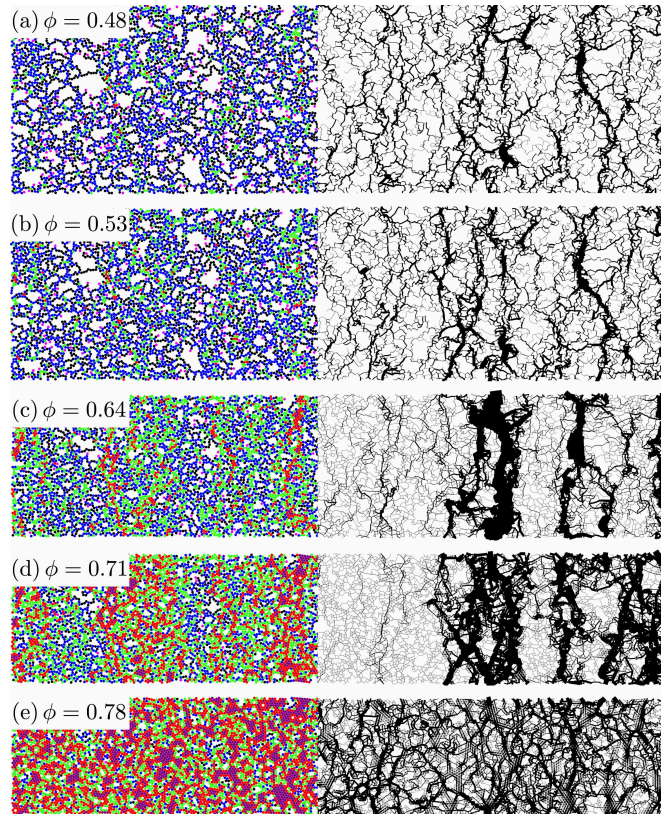}
  \caption{\label{fig:fig2-ParticleContactNum} Evolution of typical particle network for Bond 4 system, illustrated in terms of (left) particle contact number $z$ ($z=1$: pink, $z=2$: blue, $z=3$: green, $z=4$: red, $z=5$: magenta) and (right) force chains (line thickness scales logarithmically with force)}
\end{figure}

In order to probe different consolidation pathways from dilute to close-packing, we consider four different bond types as characterised by the elastic moduli $k_N$, $k_S$, $k_R$ in Table~\ref{tab:table1}, with all other model parameters (see~\cite{SM:2000}) fixed. 
These bond types are chosen to span a wide range of consolidation pathways due to the differing magnitudes of elastic moduli, even though some cases (Bond 2, Bond 3) may not be physically realistic. We note that normal forces typically dominate over sliding friction which in turn dominate rolling friction for typical colloidal particles~\cite{Fuchs:2014aa}, i.e. $k_N \gg k_S \gg k_R$. 

\begin{table}[h]
\caption{\label{tab:table1}%
Elastic moduli for Bond 1-4 systems.}
\begin{tabular}{ccccc}
\hline
moduli & Bond 1 & Bond 2 & Bond 3 & Bond 4\\
\hline
$k_N$ & 1 & 1 & 1 & 100\\
$k_S$ & 1 & 1 & 100 & 1\\
$k_R$ & 1 & 100 & 1 & 1\\
\hline
\end{tabular}
\end{table}

For each bond type, 100 realisations of particle networks consisting of $N=11140$ unit radius particles 
are constructed via a random fractal aggregation model~\cite{Seto:2013aa} in the square domain $(x,y)$=500$\times$500  (corresponding to a solids areal fraction $\phi\approx0.14$). Quasi-static consolidation is then performed via a series of strain steps along the vertical $y$-axis (up to $\phi_{\text{max}}\approx 0.78$), 
and the network is allowed to relax to equilibrium 
after each strain step via a Langevin equation, see \cite{SM:2000} for details. 
Consolidation of a typical network for the Bond 4 system via this method is shown in Fig.~\ref{fig:fig2-ParticleContactNum}, where the development of anisotropic particle and force distributions are readily apparent.

\begin{figure}[h]
\includegraphics[width=1\columnwidth]{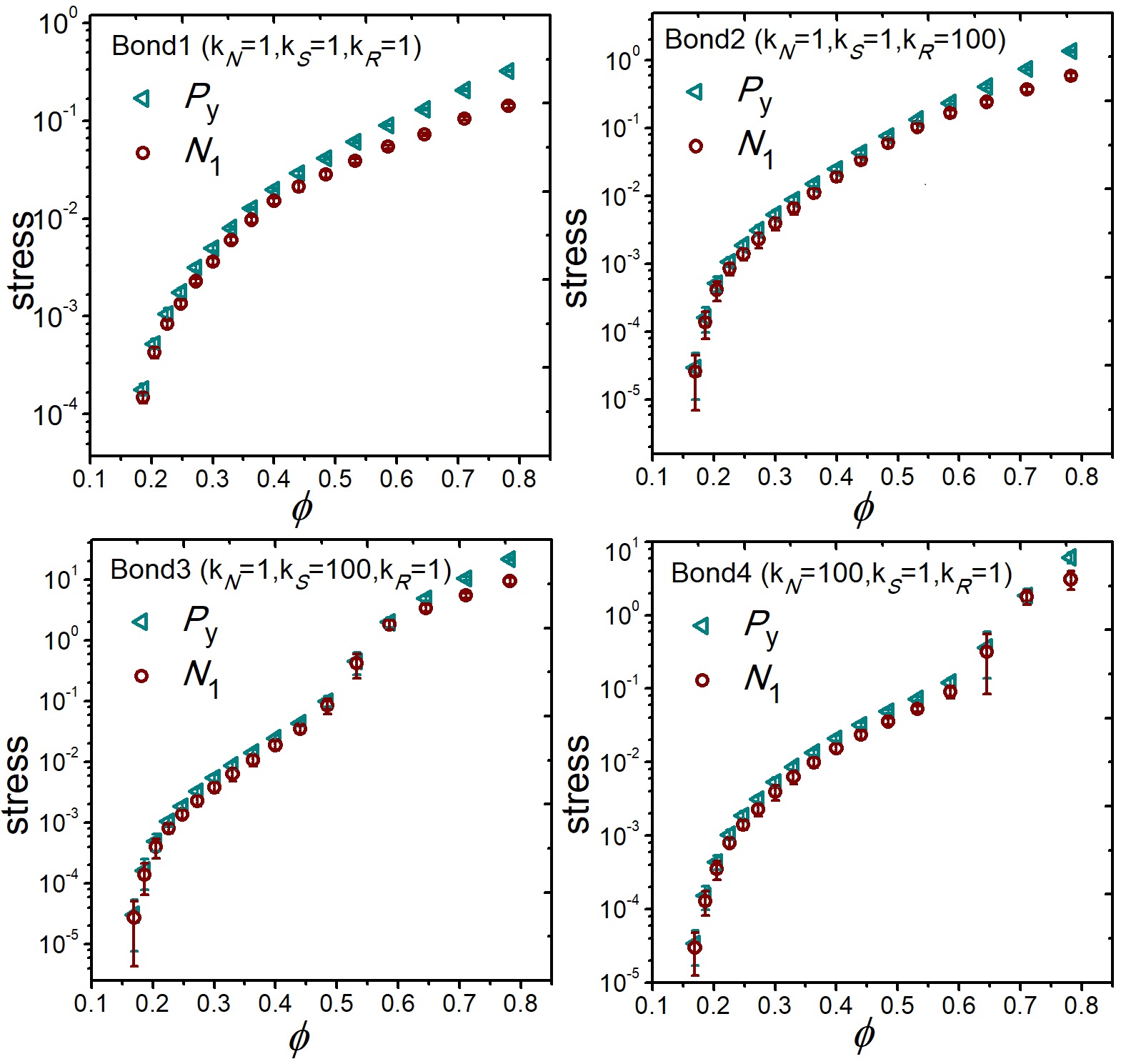}
\caption{\label{fig:fig4-NorStress} Evolution of compressive stress $P_y(\phi)$ and normal stress difference $N_1(\phi)$ for Bonds 1-4.}
\end{figure}

\begin{figure}[h]
\includegraphics[width=1\columnwidth]{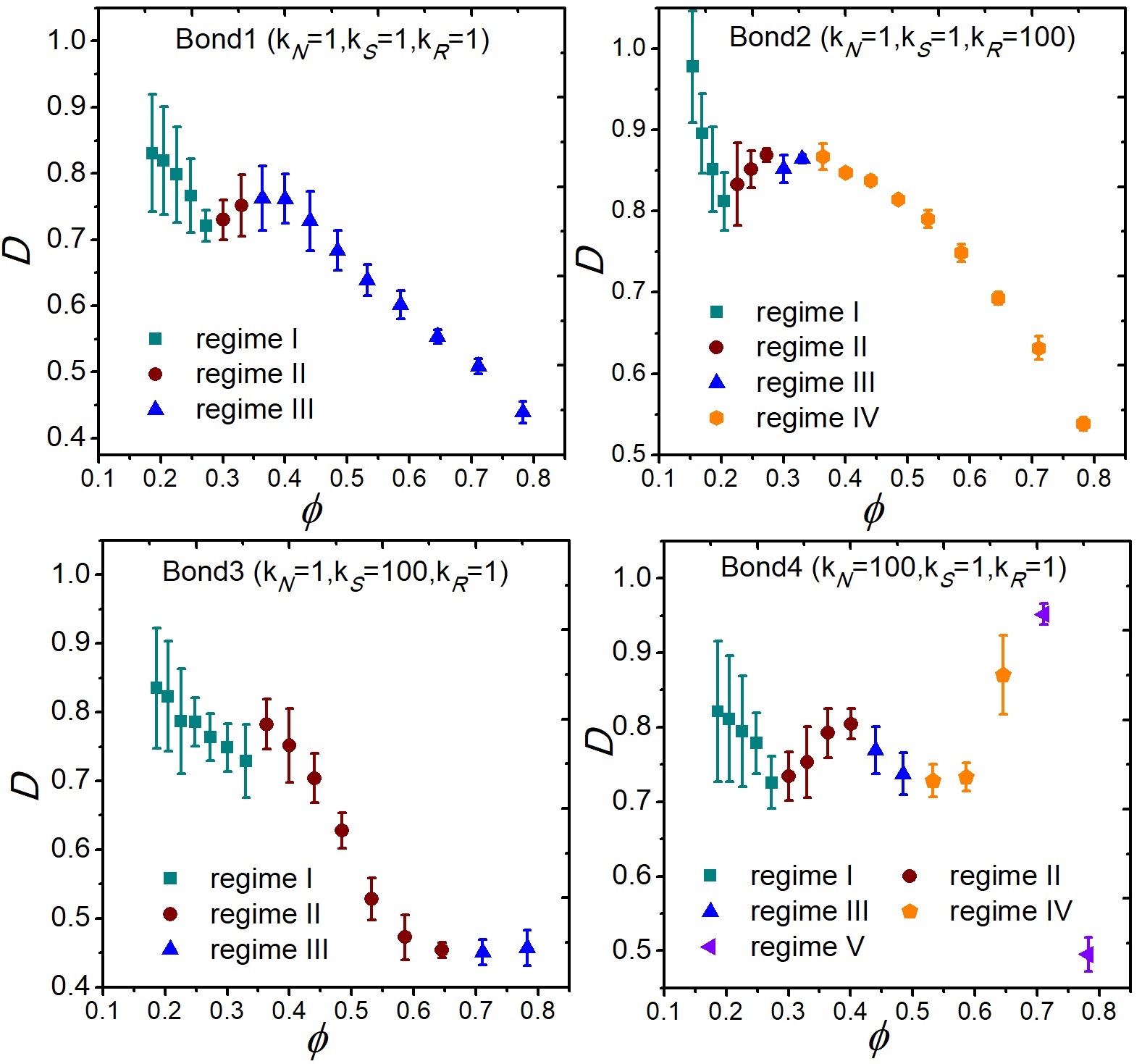}
\caption{\label{fig:fig5-Ratio}Ratio of normal stresses difference to compressive stresses for Bond 1-4.}
\end{figure}

Evolution of the axial $\sigma_{yy}$ 
and transverse $\sigma_{xx}$ stresses for all four bond types is shown in Fig.~\ref{fig:fig4-NorStress}, which show a transition from an apparent gel point $\phi_g\approx0.178$ to power law growth in all cases. Bond 3 and 4 systems also exhibit the onset of new consolidation regimes at $\phi\approx 0.55$ and $\phi\approx 0.65$ respectively. In all cases the normal stress difference $N_1(\phi)\equiv\sigma_{yy}-\sigma_{xx}$ is a significant fraction of the compressive stress $P_y(\phi)\equiv\sigma_{yy}$, in contrast to most constitutive theories~\cite{Buscall:1987aa, Brambilla:2011aa, Miller:1996aa} that implicitly assume isotropic consolidation. Whilst these stress curves are monotonic increasing and relatively featureless, the normal stress difference ratio $\chi(\phi)\equiv N_1(\phi)/P_y(\phi)\in[0,1]$ shown in Fig.~\ref{fig:fig5-Ratio} displays much richer behaviour.

In Fig.~\ref{fig:fig5-Ratio} we observe very different consolidation pathways across the four bond types due to the different hierarchies of elastic moduli 
that govern preferential consumption of rotational and translational DOFs during consolidation. Across all of these bond types $\chi(\phi)$ exhibits distinct regimes (denoted I-V) which indicate different modes of consolidation. 
One consequence of the elasto-plastic inter-particle forces is that these critical transitions are predominantly elastic at low $\phi$ due to exhaustion of elastic deformation modes, and plastic at high $\phi$ due to rigidity percolation. 
In this study we focus primarily on the origin and nature of the plastic critical transitions during consolidation.


We postulate that the observed plastic transitions in these systems are critical jamming thresholds that correspond to the directed percolation of structures that are deemed to be rigid with respect to various combinations of normal, sliding and rolling deformations between particles, leading to multiple possible jamming thresholds. For frictional assemblies, this requires extension of classical approaches to rigidity percolation for frictionless systems~\cite{Maxwell:1864aa, Jacobs:1995aa, Jacobs:1997aa, Wyart:2008aa} to incorporate sliding and rolling friction. Following classical rigidity percolation, clusters of particles within the network are determined to be either \emph{rigid} or \emph{floppy} based on whether a force-free deformation of the cluster is possible. This is determined by the nullspace of the global stiffness matrix $\mathcal{M}$ that relates the total force/torque increment $\delta\mathbf{f}$ to the translation/rotational increment $\delta\mathbf{x}$ of the particle network as $\delta\mathbf{f}=\mathcal{M}\cdot\delta\mathbf{x}$,
see \cite{SM:2000, Lester:2018aa} for details. The global stiffness matrix $\mathcal{M}$ encodes all of the normal and frictional forces in the particle network and so depends upon the elastic moduli $k_N$, $k_S$, $k_R$ as input parameters. Given the magnitudes of the elastic moduli across the different bond types in Table~\ref{tab:table1}, we consider a hierarchy of jamming states with respect to various combinations of the deformation modes.

As particles must be bonded via normal forces for the frictional forces to act, normal forces must be present in all combinations of forces under consideration. Hence we consider jamming with respect to normal forces only (N), normal and sliding forces only (NS), normal and rolling forces only (NR) and all three modes (NSR). These states are then identified by performing a rigidity percolation analysis upon the global stiffness matrix $\mathcal{M}$ with the relevant elastic moduli set to zero; e.g. $\mathcal{M}_\text{N}=\mathcal{M}|_{k_S=k_R=0}$, $\mathcal{M}_\text{NS}=\mathcal{M}|_{k_R=0}$, and the corresponding critical solids concentrations are respectively denoted $\phi_\text{N}$, $\phi_\text{NS}$. When all three forces are acting, the number of constraints (3) upon particle motion is equal to the number of DOFs (2 translational and 1 rotational), hence NSR rigidity corresponds directly to connectivity between particles, and the first jamming state (critical transition) corresponds to the onset of gelation, given by the gel point $\phi_g$. Lester et al~\cite{Lester:2018aa} show that in the absence of sliding friction, rolling and translational deformations completely decouple and so assembly rigidity with respect to NR and N are identical, i.e. rolling friction does not contribute to rigidity of the particle assembly. 
Given these constraints, the hierarchy of jamming transitions as a function of solids concentration is then ordered as $\phi_g=\phi_\text{NSR}\leqslant\phi_\text{NS}\leqslant\phi_\text{NR}=\phi_\text{N}$,
irrespective of the magnitude of the elastic moduli. Furthermore, anisotropy of the uniaxial consolidation (as illustrated by the particle clusters and force chains in Fig.~\ref{fig:fig2-ParticleContactNum} leads to \emph{directed} jamming in the $x$- and $y$-directions, where due to anisotropy the hierarchy of jamming transitions follows $\phi^{(y)}\leqslant\phi^{(x)}$ for all combinations of N, S, R as
\begin{equation}
\phi_g=\phi_\text{NSR}^{(y)}\leqslant\phi_\text{NSR}^{(x)}\leqslant\phi_\text{NS}^{(y)}\leqslant\phi_\text{NS}^{(x)}\leqslant\phi_\text{N}^{(y)}\leqslant\phi_\text{N}^{(x)}\leqslant\phi_\text{cp}
\label{eqn:jamming_hierarchy}
\end{equation}          

\begin{figure}[h]
\includegraphics[width=1\columnwidth]{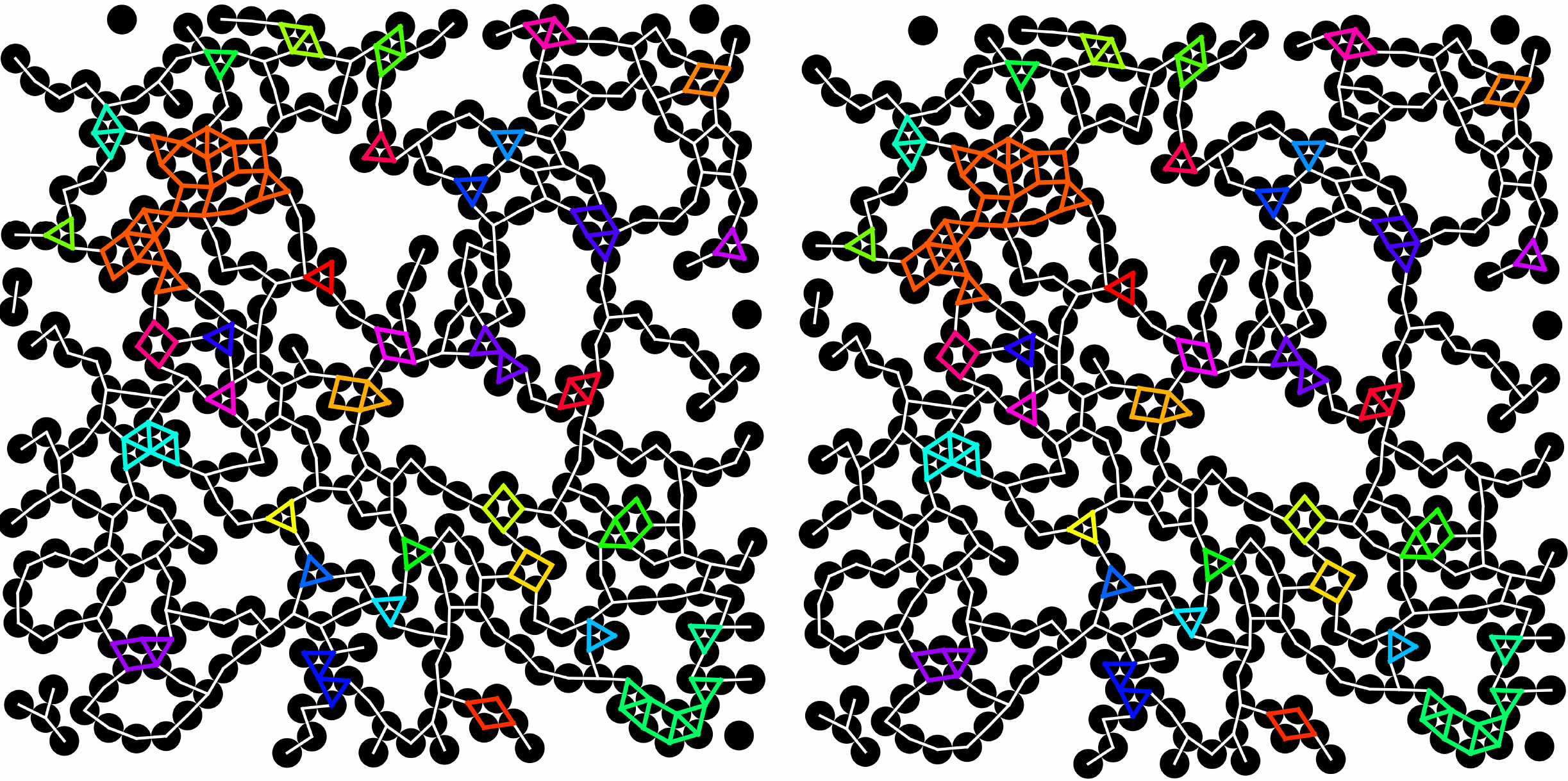}
\caption{\label{fig:rigidity}Distinct (coloured) clusters of rigid structures for a typical colloidal gel particle assembly with respect to (left) N, (right) NS rigidity. NR rigidity is equivalent to N rigidity, and NSR rigidity corresponds to the connectivity network (white lines).}
\end{figure}

As determination of the null modes of $\mathcal{M}$ is ill-posed for large systems ($N\gtrsim 100$) via continuous methods, a robust and efficient \emph{pebble game}~\cite{Jacobs:1995aa, Jacobs:1997aa} algorithm based on graph theory has been developed for frictionless systems. Henkes~\cite{Henkes:2016aa} propose that  the $(3,3)$ variant of \emph{generalised} $(k,l)$ pebble games~\cite{Lee:2008aa} (where $k$ is the DOFs per particle and $l\in[0,2k)$ is the number of trivial global motions) solve the frictional rigidity problem. However, this method fails to identify a 4-cycle of particles as being rigid with respect to translation.
Lester et al~\cite{Lester:2018aa} show that a merger of (3,3) and (3,4) pebble games correctly and robustly identifies NS rigid/floppy structures in 2D saturated 
frictional systems, and the original $(2,3)$-pebble game ~\cite{Jacobs:1995aa, Jacobs:1997aa} determines frictionless (N) rigidity. Fig.~\ref{fig:rigidity} illustrates the various rigid structures (N, NS, NR, NSR) for a typical particle assembly 

Due to the redundancies associated with NSR and NR rigidity, we apply the frictional pebble game algorithm to Bonds 1-4 (see \cite{SM:2000} for details) for the NS and N critical transitions and find (see Fig.~\ref{fig:fig5-Ratio}) that the regime II/III transition in Bond 3 coincides with simultaneous NS-$x$ and NS-$y$ rigidity percolation. Similarly for Bond 4 we observe that the regime III/IV transition coincides with NS-$x$ rigidity, regime IV/V transition coincides with simultaneous NS-$y$ rigidity and N-$x$ rigidity, and the regime V/VI transition coincides with N-$y$ rigidity. Conversely, neither NS nor N transitions are observed for Bonds 1 and 2 for reasons that shall be explained below. 

\begin{figure}[h]
\includegraphics[width=1\columnwidth]{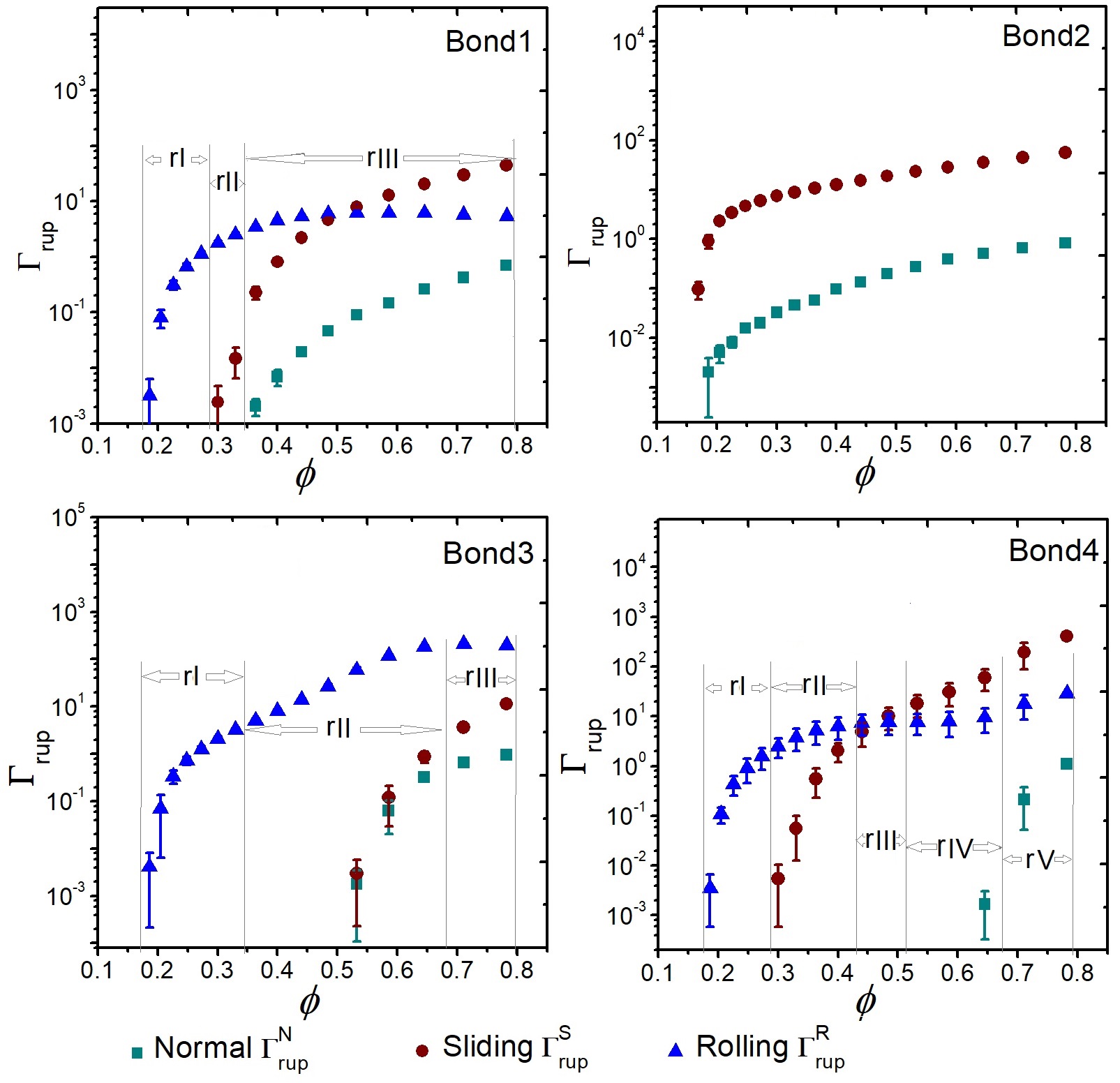}
\caption{\label{fig:fig5-BondBreakage}Bond rupture rate of Bond 1-4. 'rI' means 'regime I' which corresponds to Fig.~\ref{fig:fig5-Ratio}. Analogically 'rII','rIII','rIV' and 'rV' mean 'regime II','regime III','regime IV','regime V'}
\end{figure}

The observation that directed frictional rigidity percolation punctuates the various regimes during consolidation of strong colloidal gels confirms our hypothesis that hierarchical jamming transitions govern deformation of frictional particle assemblies, and provide the link between inter-particle forces and collective behaviour of the assembly. Here, strong preferential deformation of the particle network due to large differences in the elastic moduli across the various bond types typify each consolidation regime. As consolidation proceeds, the DOFs associated with each preferential deformation mode are exhausted, leading to a ``crisis'' in the network that is indicated by a critical jamming transition. The most basic jamming transition occurs at the onset of gelation for all (Bond 1-4) systems, where both vertical and horizontal connectivity (and hence NSR rigidity) occur simultaneously at $\phi_g=\phi_\text{NSR}^{(y)}=\phi_\text{NSR}^{(x)}\approx 0.17$ (see \cite{SM:2000}).

The next jamming transition (NS, according to (\ref{eqn:jamming_hierarchy})) is most striking in the Bond 4 system, where dense, vertical particle columns evolve that localise force transmission (see Fig.~\ref{fig:fig2-ParticleContactNum}) and correspond to NS-rigid structures (see Fig.~\ref{fig:rigidity}) and lead to NS-$y$ rigidity percolation at $\phi_\text{NS}^{(y)}\approx0.525$ as the rolling DOFs are exhausted. This transition marks the onset of buckling of these vertical force chains, the onset of breakage of sliding bonds (see Fig.~\ref{fig:fig5-BondBreakage}), and a plateau in $\chi(\phi)$ corresponding to a reduction in anisotropy of the network, which eventually leads to NS-$x$ rigidity percolation at $\phi_\text{NS}^{(x)}\approx0.635$ as the buckling NS-rigid chains coalesce in the horizontal direction.

This transition marks the onset of a rapid rise in $\chi(\phi)$ as the network densifies once again in a highly anisotropic fashion, now characterised by dense, N-rigid vertical force chains as the sliding DOFs are exhausted in the system, leading to N-$y$ rigidity percolation at $\phi_\text{N}^{(y)}\approx0.661$. After this transition, the breakage of normal bonds occurs for the first time in the system as again the N-rigid vertical force chains buckle, leading to a convex evolution of $\chi(\phi)$ until the buckled N-rigid chains coalesce in the $x$-direction, leading to N-$x$ rigidity percolation at $\phi_\text{N}^{(x)}\approx0.735$. This final jamming transition is associated with a steep decrease in $\chi(\phi)$ as the particle microstructure and associated force chains become increasingly isotropic as the remaining normal force DOFs in the system are consumed, ultimately leading to a halt in consolidation (and divergence of compressive stress) at the close-packing limit $\phi_\text{cp}\approx 0.84$. Note that whilst the network is spatially isotropic at close-packing, $\chi(\phi_\text{cp})$ is still non-zero here due to residual plastic stresses and hysteresis artefacts from the uniaxial consolidation process.

Similarly, for Bond 3 system the onset of NS-$y$ rigidity at $\phi_\text{NS}^{(y)}\approx0.648$ marks the onset of both sliding and normal bond breakage and a plateau of $\chi(\phi)$ as vertical force chains that are NS-rigid buckle and eventually coalesce. Whilst the jamming hierarchy (\ref{eqn:jamming_hierarchy}) still holds, the remaining transitions do not occur until the system is very near to the close-packing limit $\phi_\text{cp}$. This is due to the change in elastic moduli (specifically $k_S\gg k_N$), meaning that sliding deformations are energetically unfavourable, hence the hierarchical jamming transitions (\ref{eqn:jamming_hierarchy}) do not occur until very large stresses are applied. For the Bond 1 and 2 systems, this delay is even more pronounced as for all of these cases $k_R$ is maximal and $k_S$ is minimal, hence both the NS and N transitions are energetically unfavourable and so occur near the close-packing limit. Indeed, rolling bond breakages never occur for the Bond 2 system as sliding deformation is always more energetically favourable. Consolidation of these systems is then governed solely by elastic deformations, as is illustrated in \cite{SM:2000}. It is important to note that the typical hierarchy of frictional and normal forces ($k_N\gg k_S\gg k_R$) observed in granular systems~\cite{Fuchs:2014aa} means that the jamming hierarchy (\ref{eqn:jamming_hierarchy}) is most likely to be spread out over the consolidation path (i.e. similar to Bond 4 system) rather than confined near to the close-packing limit (i.e.similar to Bond 1 and 2 systems).

These observations show that hierarchical jamming governs the consolidation of strong colloidal gels, where critical
jamming transitions correspond to frictional and frictionless directed rigidity percolation. This hierarchy of jamming states (\ref{eqn:jamming_hierarchy}) arises from purely kinematic constraints and so is independent of the particle-scale forces. These forces however inform how the jamming hierarchy manifests (i.e. the spacings between critical transitions) and the maximum-likelihood consolidation pathway of the system. These concepts apply to general deformation of a broad range of frictional disordered solids, and provide deep insights into the evolution of these complex systems. Future studies include extension of these ideas to 3D (including torsional deformation) and aspherical particles, and classification of these jamming transitions and examination of universality of the associated scaling exponents.

\providecommand{\noopsort}[1]{}\providecommand{\singleletter}[1]{#1}%

\end{document}